\documentclass[a4paper, 10pt, conference]{ieeeconf}      

\IEEEoverridecommandlockouts                              

\usepackage{graphicx}
\usepackage{algorithm}
\usepackage{algorithmic}
\usepackage{hyperref}
\usepackage{xcolor}
\usepackage{subcaption}

\title{\LARGE \bf
Underground Multi-robot Systems at Work: a revolution in mining
}

\author{Victor V. Puche\textsuperscript{*}\textsuperscript{1}, Kashish Verma\textsuperscript{*}\textsuperscript{1}, and Matteo Fumagalli\textsuperscript{1}
\thanks{*These authors contributed equally to this work.}
\thanks{\textsuperscript{1}The authors are with the Department of Electrical and Photonics Engineering of the Technical University of Denmark.
        {\tt\small \{vvipu, s230015, mafum\}@dtu.dk}
}
}

\begin{document}

\maketitle
\thispagestyle{empty}
\pagestyle{empty}

\begin{abstract}
The growing global demand for critical raw materials (CRMs) has highlighted the need to access difficult and hazardous environments such as abandoned underground mines. These sites pose significant challenges for conventional machinery and human operators due to confined spaces, structural instability, and lack of infrastructure. To address this, we propose a modular multi-robot system designed for autonomous operation in such environments, enabling sequential mineral extraction tasks. Unlike existing work that focuses primarily on mapping and inspection through global behavior or central control, our approach incorporates physical interaction capabilities using specialized robots coordinated through local high-level behavior control. Our proposed system utilizes Hierarchical Finite State Machine (HFSM) behaviors to structure complex task execution across heterogeneous robotic platforms. Each robot has its own HFSM behavior to perform sequential autonomy while maintaining overall system coordination, achieved by triggering behavior execution through inter-robot communication. This architecture effectively integrates software and hardware components to support collaborative, task-driven multi-robot operation in confined underground environments.

\end{abstract}

\section{INTRODUCTION}
The accelerating global demand for critical raw materials (CRMs) - such as rare earth elements, lithium, and cobalt - for clean energy, digital technologies, and strategic manufacturing has intensified the urgency to secure reliable, and local sources within Europe \cite{c3, c4}. Deep and abandoned underground mines represent a promising but highly challenging opportunity due to their confined dimensions, structural risks, and lack of infrastructure. Human access is often unsafe or impossible, and traditional machinery lacks the autonomy and agility required to operate in such environments. Addressing these challenges requires the development of modular multi-robot systems capable of operating autonomously in confined, infrastructure-less underground environments to perform a wide range of tasks, including exploration, maintenance, and drilling. By distributing responsibilities across specialized robots and enabling onboard decision-making, these systems could improve safety, scalability, and mission robustness in high-risk settings.

Recent research initiatives have demonstrated the potential of robotic systems in addressing these challenges. However, most efforts have focused on mapping and search-and-rescue operations, with limited attention given to physical interaction tasks such as drilling. Various robot locomotion platforms have been explored for underground environments. The Groundhog project \cite{c6} was among the first to deploy underground robots, using a compact four-wheeled Ackermann-steered vehicle to generate 2D maps in partially collapsed mines. When terrain becomes too complex for ground vehicles, Micro Aerial Vehicles (MAVs) offer a promising alternative. Equipped with onboard sensors, MAVs can inspect inaccessible or low-visibility areas \cite{c7}. Multi-robot systems further enhance operational capabilities. For example, in \cite{c8}, a legged robot is extended to carry an aerial platform, enabling rapid deployment in search-and-rescue scenarios.

Some efforts have also addressed underground manipulation tasks. In \cite{c9}, a mobile manipulator was developed to interact with Smart Sensor Boxes (SSBs) that monitor environmental parameters. The ARIDuA project explores how such a system could install, rearrange, or remove SSBs in situ.

More recently, the H2020 ROBOMINERS project \cite{c5} introduced RM1, a bio-inspired full-scale robot designed for operation in confined underground spaces. RM1 integrates mapping and ore detection sensors, a drilling tool for rock excavation, and a mechanism for material transport. The robot features a compact form (800 mm diameter, 1500 mm length, ~1500 kg), but relies entirely on water hydraulics and tethered operation, limiting its autonomy. Although RM1 proves the viability of robotic deployment in confined spaces, its reliance on water hydraulics and tethered operation highlights the need for modular, untethered, multirobot systems coordinated through robust high-level control. 

High-level control plays a crucial role in enabling complex modular multi-robot systems to become a reality. Two widely used approaches for deploying high-level control are Hierarchical Finite State Machines (HFSMs) \cite{c11} and Behavior Trees (BTs) \cite{c12}. However, each approach has its own advantages and limitations. For the DARPA Robotics Challenge (DRC) from 2012 to 2015, which was aimed to develop semi-autonomous ground robots that perform complex tasks in dangerous, degraded, human-engineered environments, Team ViGIR developed a Collaborative Autonomy system, which is known as Flexible Behavior Engine or FlexBE \cite{c11}. FlexBE facilitates the ease of designing, developing and deploying cyclic and complex behaviors based on HFSMs, whereas implementing cyclic behaviors using Behavior Trees (BTs) is less intuitive with existing tools such as BehaviorTree.CPP \footnote{See \href{https://github.com/BehaviorTree/BehaviorTree.CPP}{https://github.com/BehaviorTree/BehaviorTree.CPP}} and py\_trees\_ros \footnote{See \href{https://github.com/splintered-reality/py\_trees\_ros}{https://github.com/splintered-reality/py\_trees\_ros}}.

Moreover, the existing multirobot systems utilize a single global behavior to execute the missions which also require persistent connectivity with the central behavior control. To this end, we propose a system, utilizing multiple FlexBE HSFMs behavior, in which each robot executes its respective HFSM behavior upon receiving specific triggering communication messages. This ensures a sequential pipeline of tasks that can be performed by more than one robot without persistent connectivity to any central machine.

\section{APPROACH}

\subsection{Requirements and boundaries}

Autonomous mineral exploration and drilling in deep and abandoned underground environments imposes a set of physical and environmental boundaries that fundamentally shape the system design. Mine tunnels typically offer limited space, often constrained to cross-sections no longer than 2m x 2m, which limits the size, maneuverability, and actuation capabilities of each robotic unit. In addition, such environments lack both GPS signals and wireless connectivity, requiring all sensing, navigation, and coordination to be managed entirely onboard. Operational reliability is further challenged by factor such as low visibility, moisture, and uneven terrain. Energy autonomy is also critical due to the lack of power infrastructure, necessitating onboard power management and resupply strategies to maintain functionality over time. 

In response to these constraints, the robotic system must meet several essential requirements. It must operate fully autonomously in unstructured and GPS-denied settings, relying solely on onboard sensors.  The architecture must support a heterogeneous, modular fleet, in which each robot is specialized for tasks such as exploration, manipulation, deployment, or drilling, while remaining adaptable to evolving mission needs. Multi-robot collaboration is crucial: robots must coordinate their roles and synchronize mission phases without centralized control, adapting in real time to dynamic events such as unexpected obstacles or partial failures. Furthermore, the system must be fault-tolerant able to detect issues such as anchoring instability or drilling misalignment, recover autonomously, and continue mission execution safely and efficiently.

Based on the explained requirements, and continuing with the mission conceptualization done in \cite{c2}, the following robotic agents are required to perform the mission. 
\begin{itemize}
    \item Explorer robot: A ground vehicle with exploration and navigation capabilities. Its role includes exploration, mapping, and the detection of potential drilling sites.
    \item Deployer robot: A ground vehicle with navigation and manipulation capabilities. Its roles include 3D environment surface analysis, and manipulation tasks such as placing and servicing the drilling tool.
    \item Supplier robot: A logistics agent designed to transport essential resources like power units, drilling consumables, water, etc., supporting the drilling operation.
    \item Stinger robot: A compact drilling agent with a deployable anchoring system and an integrated drilling unit. It is capable of self-anchoring, and performing multi-hole drilling operations. 
\end{itemize}

\begin{figure}
    \begin{minipage}{0.5\textwidth}
      \centering
      \includegraphics[width=\textwidth]{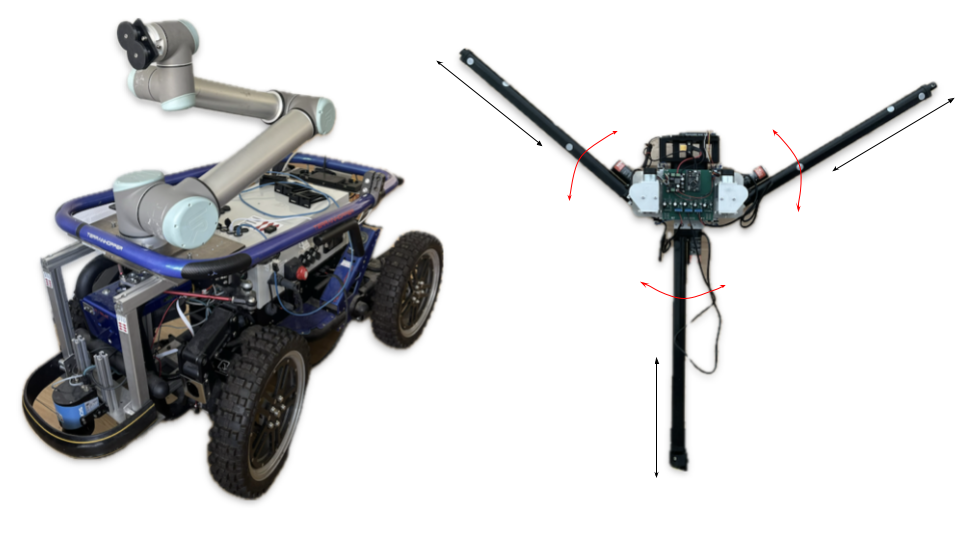}
    \end{minipage}
  \caption{Deployer and Stinger Robot}
  \label{fig:robots}
\end{figure}

\subsection{Concept of Operation (mission)}\label{Concept of operation}
The autonomous exploration and drilling mission begins with one or more Explorer robots tasked with navigating uncharted sections of the mine to generate a global map of the environment and identify potential drilling sites based on ore detection. This initial exploration phase produces the spatial and semantic information required to plan and guide the rest of the robotic fleet. Once suitable drilling candidate locations are identified, the Deployer, Supplier, and Stinger robots navigate to the selected site using the pre-acquired map.

This paper focuses on the drilling phase of the mission, which begins once the robotic fleet (Deployer and Stinger robot as shown in fig.\ref{fig:robots}) has reached the designated drilling area. Upon arrival, the Deployer robot uses its onboard sensors to analyze the local terrain and determine both the most suitable drilling site and the optimal deployment configuration for the Stinger robot. It then retrieves the stinger robot and, using its onboard manipulator, precisely places it at the selected location. The Stinger robot anchors itself to the tunnel surface and autonomously executes the drilling operation. Throughout this process, the Supplier robot, which is under development, provides support when necessary, for instance, by delivering power or replacement tools to maintain continuous operation. Once the drilling task is completed, the Deployer may reposition the Stinger robot to a new site and repeat the process. This collaborative workflow enables safe and efficient autonomous drilling in confined, infrastructure-less underground environments.

\section{SYSTEM DESIGN AND VERIFICATION}
Based on the requirements outlined earlier and the mission conceptualization, we have identified a set of robotic agents, each with distinct roles and capabilities, necessary to carry out the autonomous drilling mission. These capabilities must be translated into specific hardware solutions that can operate reliably in a constrained and challenging underground environment. 

\subsection{Breakdown of hardware elements and controls}
\subsubsection{Deployer Robot}
The Deployer robot is built on a modified Terrain Hopper Overlander Sprint, a compact two-wheel-drive electric vehicle designed for off-road mobility in uneven environments. It can handle slopes of up to 15° uphill and 30° downhill, and supports a maximum payload of 104 kg to transport mission equipment. Its narrow width of 850 mm allows it to navigate within confined 2 m × 2 m mine tunnels, while waterproof motors ensure reliable performance in humid underground conditions. The platform is powered by a 24 V lithium battery, offering a driving range of up to 45 km, suitable for extended missions. The seat has been removed and the chassis has been modified to accommodate the onboard hardware.

Mobility and steering are controlled via two Roboteq MDC2460 motor controllers (one per axle) and an additional Roboteq MDC1460 for steering, all integrated on a CANopen bus running at 250 kbit/s. These controllers interface with the main onboard computer through a USB-to-CAN adapter. Rear wheel motion feedback is provided by AS5057 magnetic encoders mounted on the motors, while steering feedback is obtained via an AS5045 encoder read by a Teensy 3.1 microcontroller. The Teensy also reads analog inputs from the battery and a steering potentiometer, using voltage dividers for compatibility.

Mounted on the vehicle is a 6-DoF UR10 industrial robotic arm, powered via a DC-DC converter connected to the Terrain Hopper’s battery. With a 1.3 m reach and 10 kg payload capacity, it enables manipulation tasks such as deploying the Stinger robot and handling payloads within confined spaces. A WiFi router, TP-Link Pharos CPE210, is installed onboard to establish a local network that connects all robotic agents during the mission.

\subsubsection{Stinger Robot}
Stinger robot is a three-deployable leg robot which has a self-stabilizing bracing system, see Fig.\ref{fig:robots}. The black line shows the movement of linear actuator and red lines shows the rotational movement of legs. The Onboard latte panda is utilized for running the software architecture explained in this paper. This paper focuses on system-level integration; for detailed technical background of the hardware elements and its control, refer to Stinger Robot paper.

\subsection{Simulation} 
A simulation environment has been set up to develop and test the algorithms and behaviors of the system. The Gazebo simulator \cite{c10} was chosen for its seamless integration with the ROS2 autonomy stack and its rich library of sensors and actuators, which allows realistic emulation of robotic capabilities. The robotic agents involved in the mission have been included by developing or reusing existing models defined using the Unified Robot Description Format (URDF).

\section{SOFTWARE ARCHITECTURE AND MISSION MODULES}
\label{sec:soft_archi}
\begin{figure*}[htbp]
  \centering
  \fbox{%
    \begin{minipage}{0.95\textwidth}
      \centering
      \includegraphics[width=\textwidth]{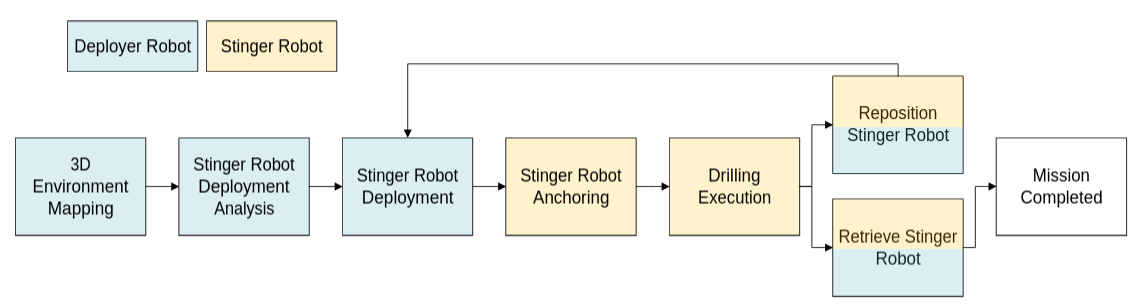}
    \end{minipage}
  }
  \caption{Mission software modules and their interactions during the drilling phase.}
  \label{fig:mission_modules}
\end{figure*}

To implement the concept of operation described above, the high-level autonomous drilling mission must be broken down into discrete software modules that correspond to specific robot behaviors. Figure \ref{fig:mission_modules} illustrates the mission structure and its submodules, showing the tasks assigned to the Deployer robot (blue modules) and Stinger robot (yellow modules). Combined-color modules indicate collaborative actions between both robots. Despite being executed sequentially, the task qualifies as a collaborative multi-robot operation due to the shared goal and interdependence of actions. The outcome is not achievable by either robot alone, and the coordination reflects a functional division of labor typical in collaborative systems \cite{corke2013}–\cite{kalra2005}.

We utilized a high-level behavioral control strategy, HFSMs, to deploy the robots. The states were developed with a combination of substates for complex tasks/actions. To enable coordination between robots, the behavior control of a robot triggers the behavior control of another robot. This establishes a sequential control of the whole mission performed by multi-robots. However, concurrent tasks in which two robots perform actions together to achieve a common goal concurrently can be viable with some additions to this approach. 

The pipeline for triggering the execution of a behavior on another robot involves one robot completing its current behavior and then publishing an ROS 2 topic message, as shown in algorithm \ref{alg:trigger_algo}. This message serves as a trigger to initiate the behavior of the target robot that has a static IP denoted by IP$(i+1)$. This approach facilitates the coordinated and sequential execution of mission modules among the robots.

\begin{algorithm}
\caption{Coordinated Behavior Triggering in Multi-Robot System}
\begin{algorithmic}[0]
\label{alg:trigger_algo}
\REQUIRE Set of robots $R = \{R_1, R_2, ..., R_n\}$ with assigned behaviors $B_i$
\STATE Initialize ROS 2 system and start behavior for all robots
\FORALL{robot $R_i \in R$}
    \STATE Waiting FSM until trigger message is received on topic \texttt{trigger\_R\_i}
    \STATE Execute assigned behavior $B_i$
    \STATE Upon completion \WHILE{ping\_IP$_{i+1}$=\FALSE}
        \STATE Wait
    \ENDWHILE
    \IF{ping\_IP$_{i+1}$ = \TRUE} 
        \STATE Publish trigger message to topic \texttt{trigger\_R$_{i+1}$} 
    \ENDIF 
\ENDFOR
\end{algorithmic}
\end{algorithm}

\section{IMPLEMENTATION AND SYSTEM INTEGRATION}
Both the Deployer and the Stinger robots are equipped with onboard computers running Ubuntu 22.04 and ROS2 Humble, which serve as the middleware for system-level integration and control. To enhance system scalability, improve fault isolation, and manage communication triggering more effectively, the two robots were deployed using different ROS 2 domain IDs. Domain IDs provide a deeper level of communication isolation at the middleware level. This ensures that message traffic from one robot does not interfere with another, offering an added layer of modularity, security, and fault containment — particularly important in distributed or physically separated multi-robot systems. Additionally, this separation was necessary as the FlexBE Behavior Engine cannot be instantiated multiple times within the same domain ID. In order to facilitate the triggering communication messages, they are transferred across different domain IDs through \texttt{ros2/domain\_bridge} \footnote{See \href{https://github.com/ros2/domain_bridge}{https://github.com/ros2/domain\_bridge}}.

HFSMs were created using FlexBE, a tool for developing and executing HFSMs. The graphical user interface of FlexBE helps to easily develop, execute, and monitor the whole behavior. Both robots were deployed with their respective behavior, i.e. a HFSM behavior. However, the completion of Deployer's behavior triggers the execution of Stinger's behavior as explained in Section \ref{sec:soft_archi}.

\subsection{Deployer robot}
The Deployer robot integrates two main components, the Terrain Hopper mobile base and the UR10 manipulator, into a unified robotic platform controlled by a single onboard computer.

For the Terrain Hopper, a custom ROS2 node interfaces with the vehicle's hardware to manage wheel motion and steering. Low-level control is handled via the MobotWare system developed at the Technical University of Denmark \cite{c1}, specifically using its rhd module for sending steering and velocity commands. This ROS2 node subscribes to control topics for motion commands and publishes sensor feedback such as steering angle and wheel encoder data for use by other nodes in the system.

The UR10 manipulator, which lacks native ROS2 support, is integrated through a Docker container running ROS1. Within the container, the \texttt{ros-industrial-attic/ur\_modern\_driver} ROS package\footnote{See \href{https://github.com/ros-industrial-attic/ur_modern_driver}{https://github.com/ros-industrial-attic/ur\_modern\_driver}} provides control of the UR10. Communication between ROS1 and ROS2 environments is achieved using the \texttt{TommyChangUMD/ros-humble-ros1-bridge-}\\
\texttt{builder} ROS1 package\footnote{See \href{https://github.com/TommyChangUMD/ros-humble-ros1-bridge-builder}{https://github.com/TommyChangUMD/ros-humble-ros1-bridge-builder}}, allowing seamless interaction between the manipulator and the rest of the ROS2-based architecture.

\begin{figure}
  \fbox{%
    \begin{minipage}{0.45\textwidth}
      \centering
      \includegraphics[width=\textwidth]{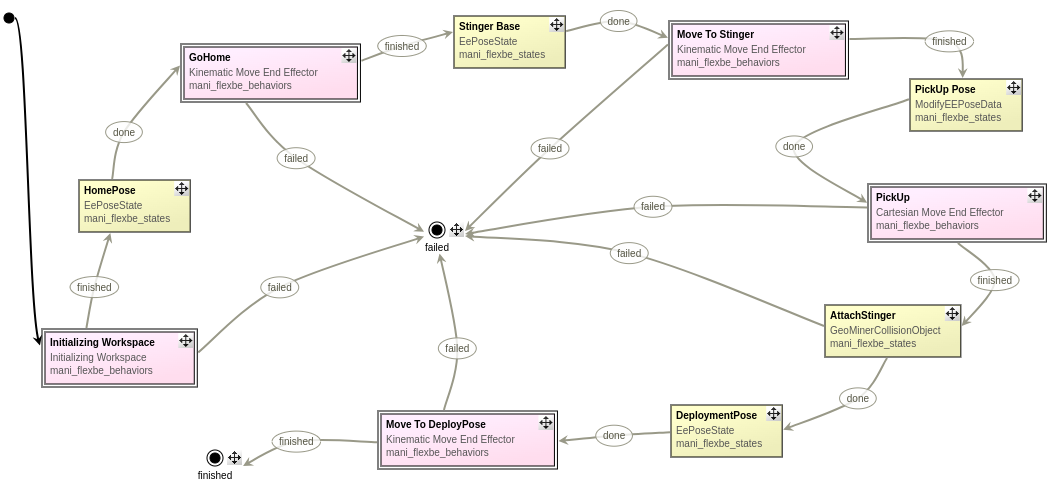}
    \end{minipage}
  }
  \caption{Deployer (UR10) Behavior to deploy Stinger robot.}
  \label{fig:ur10_b}
\end{figure}

With the ability to pick and place, the UR10 can deploy the Stinger robot. The high-level behavior was developed and executed to achieve the same, as shown in Fig. \ref{fig:ur10_b}.   

\begin{figure}
  \fbox{%
    \begin{minipage}{0.45\textwidth}
      \centering
      \includegraphics[width=\textwidth]{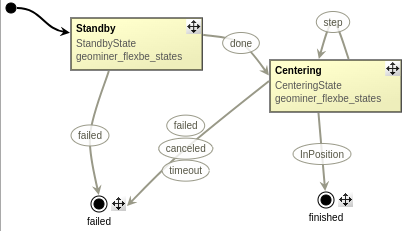}
    \end{minipage}
  }
  \caption{Stinger Behavior to deploy its anchoring legs}
  \label{fig:stinger_b}
\end{figure}

\subsection{Stinger robot}
A custom ROS2 node communicates with the Arduino to send commands to the respective rotational motors or linear actuators. The node also establishes a custom action server \texttt{move\_motor} that controls the actuators of Stinger robot. The \texttt{move\_motor} action server controls one actuator at a time for simplicity of deployment. In further developments, the complex algorithms to deploy the three stinger will be incorporated with custom action servers. 

\begin{figure*}[htbp]
    \centering
    \begin{subfigure}{0.45\textwidth}
        \centering
        \includegraphics[width=\linewidth]{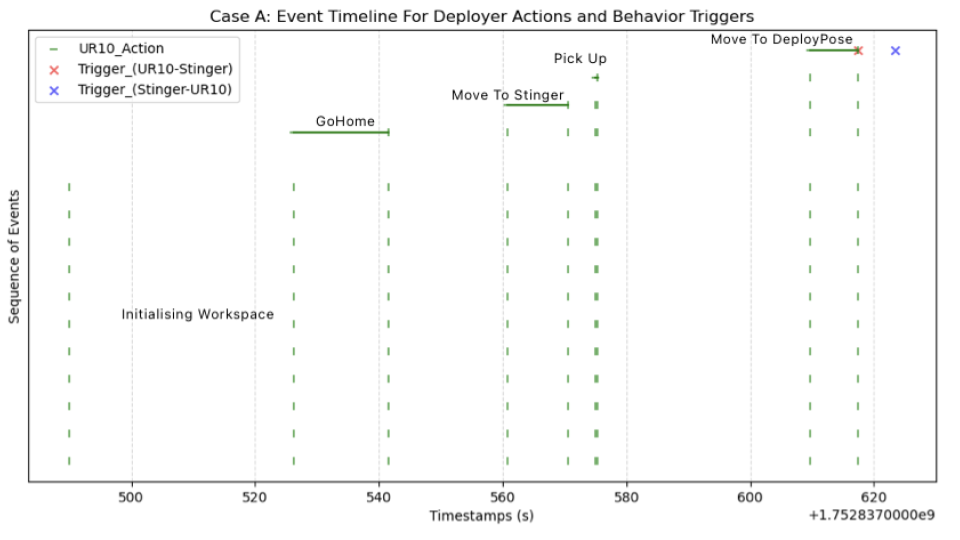}
        \label{fig:image1}
    \end{subfigure}
    \hspace{0.05\linewidth}
    \begin{subfigure}{0.45\textwidth}
        \centering
        \includegraphics[width=\linewidth]{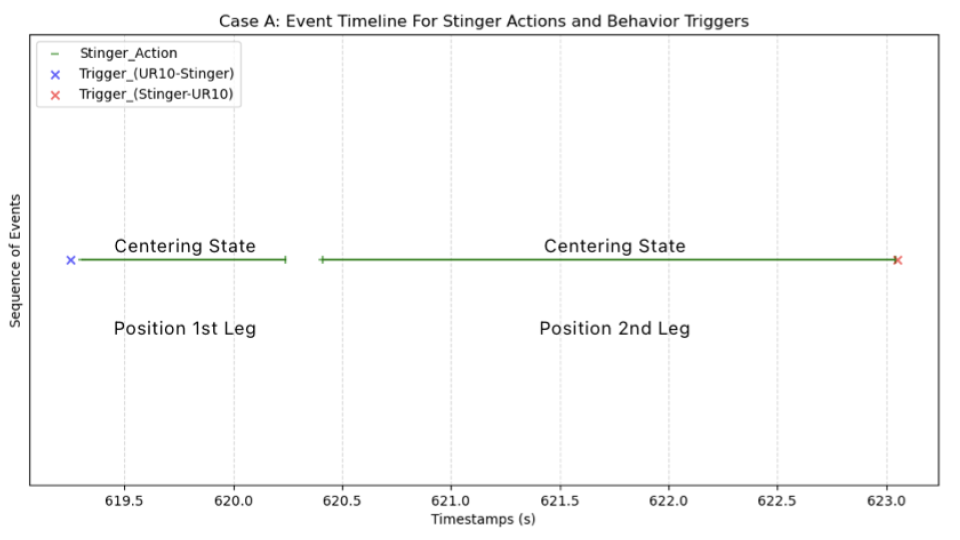}
        \label{fig:image2}
    \end{subfigure}
    \caption{Case A shows two behaviors coordinated with persistent connectivity }
    \label{fig:results_A}
\end{figure*}

\begin{figure*}[htbp]
    \centering
    \begin{subfigure}{0.45\textwidth}
        \centering
        \includegraphics[width=\linewidth]{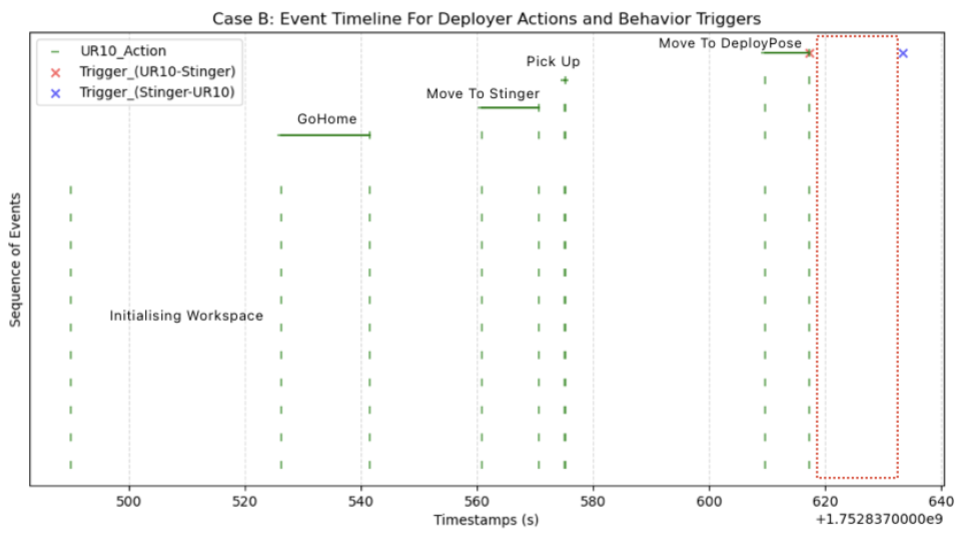}
        \label{fig:image1}
    \end{subfigure}
    \hspace{0.05\linewidth}
    \begin{subfigure}{0.45\textwidth}
        \centering
        \includegraphics[width=\linewidth]{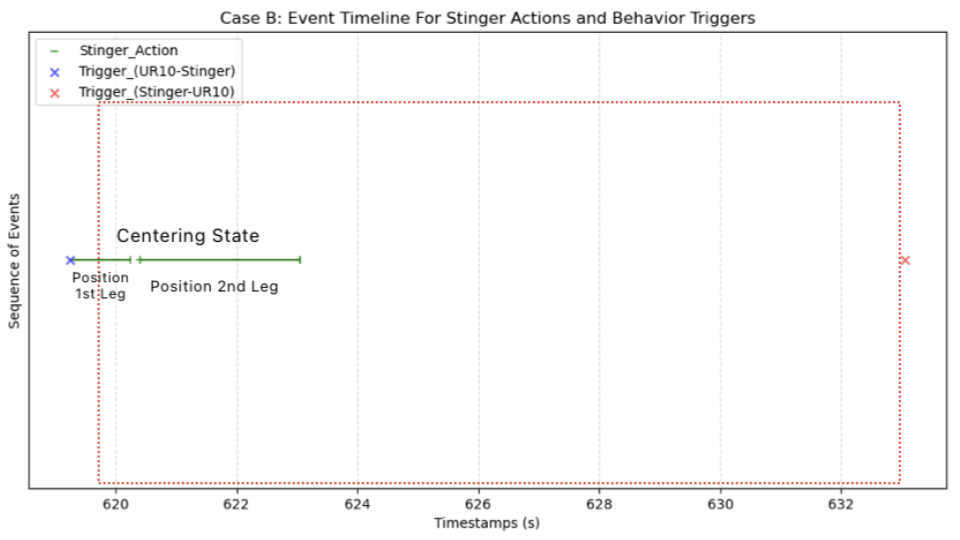}
        \label{fig:image2}
    \end{subfigure}
    \caption{Case B shows two behaviors coordinated even when the network turned off for few secs after first behavior trigger, the red dotted box represents the network off status}
    \label{fig:results_B}
\end{figure*}

The high-level behavior to control the deployment sequence of the stinger robot is shown in fig.\ref{fig:stinger_b}. The \texttt{Centering} state calls the \texttt{move\_motor} action server, and the state itself controls the side actuators one by one, e.g. left stinger (left leg) moves to its specified position, then right stinger (right leg) moves to its specified position. After reaching desired pose, \texttt{Centering} state pings the IP of Deployer to send the other trigger message.

\section{Evaluation}
The proposed software module and hardware was launched to execute the mission based on the behaviors presented in the previous section based on two cases, i.e. with persistent wireless connection and without persistent wireless connection. The objective was to record the event timeline of the respective robots for both cases as shown in fig. \ref{fig:results_A}-\ref{fig:results_B} and to also measure latency. The maximum latency was recorded as 500 ms. 

\subsubsection{Persistent Connectivity (Case A)}: The event timelines of both robots show that after picking up the Stinger robot and placing it in deployment pose, the UR10 behavior state \texttt{Move To DeployPose} sends the trigger message to the Stinger robot. Upon receiving the trigger message, the Stinger robot starts executing its own behavior, positioning two legs one by one.

\subsubsection{Without Persistent Connectivity (Case B)}: The event timeline of the Deployer robot shows that the wifi turned off after sending the trigger message, see fig.\ref{fig:results_B}. Even without connectivity, the Stinger robot continues executing its tasks in the behavior. After action completion, Stinger waits for the network connectivity to send its completion status. The moment network switches on, it send the next behavior trigger. This ensures sequential autonomy between two robots.

\section{CONCLUSIONS}
This paper presented the feasibility of the execution of multiple behaviors with integration of the hardware and software perspective of a multirobot mining system to achieve a shared objective. This system ensures the collaborative execution of the mission modules corresponding to each individual robot involved. We highlighted the necessity of proposing and implementing a multirobot system, a fleet of robots, equipped to collaborate and execute mining operations in confined and infrastructure-less environments. In the coming months, mission modules will be extended to include more complex tasks and algorithms for mining operations.

\section{Acknowledgement}
This work was prepared in the scope of PERSEPHONE project which has received funding from the European Union’s Horizon Europe Research and Innovation Programme under the Grant Agreement No.101138451.

\addtolength{\textheight}{-12cm}   





\end{document}